\documentclass[groupedaddress,
 reprint,
 amsmath,amssymb,
 aps,
 prd,
]{revtex4-1}

\usepackage{aas_macros_new}
\usepackage{graphicx}
\usepackage{dcolumn}
\usepackage{bm}
\usepackage{hyperref}
\usepackage[utf8]{inputenc}
\usepackage{dsfont}
\usepackage{color}
\usepackage{verbatim}
\usepackage{appendix}

\usepackage{mathtools}
\usepackage{physics}
\usepackage{braket}

\begin{document}

\title{Deep zoom-in simulation of a fuzzy dark matter galactic halo}

\author{Bodo Schwabe}
\email{bschwabe@unizar.es}
\affiliation{CAPA \& Departamento de F\'isica Te\'orica, Universidad de Zaragoza, 50009 Zaragoza}
\affiliation{%
 Institut f\"ur Astrophysik 
 Universit\"at G\"ottingen
}

\author{Jens C. Niemeyer}
\email{jens.niemeyer@phys.uni-goettingen.de}
\affiliation{%
 Institut f\"ur Astrophysik 
 Universit\"at G\"ottingen
}%

\date{\today}

\begin{abstract}
Fuzzy dark matter (FDM) made of ultra-light bosonic particles is a viable alternative to cold dark matter (CDM) with clearly distinguishable small-scale features in collapsed structures. On large scales, it behaves gravitationally like CDM deviating only by a cut-off in the initial power spectrum and can be studied using N-body methods. In contrast, wave interference effects near the de Broglie scale result in new phenomena unique to FDM. Interfering modes in filaments and halos yield a stochastically oscillating granular structure which condenses into solitonic cores during halo formation. Investigating these highly non-linear wave phenomena requires the spatially resolved numerical integration of the Schrödinger equation. In previous papers we introduced a hybrid zoom-in scheme that combines N-body methods to model the large-scale gravitational potential around and the mass accretion onto pre-selected halos with simulations of the Schrödinger-Poisson equation to capture wave-like effects inside these halos. In this work, we present a new, substantially improved reconstruction method for the wave function inside of previously collapsed structures. We demonstrate its capabilities with a deep zoom-in simulation of a well-studied sub-$L_\ast$-sized galactic halo from cosmological intitial conditions. With a particle mass of $m = 2.5\times 10^{-22}\,$eV and halo mass $M_{\text{vir}}=1.7\times 10^{11}\,M_{\odot}$ in a ($60$h${^{-1}}$ comoving Mpc)${}^{3}$ cosmological box, it reaches an effective resolution of 20 comoving pc. This pushes the values of $m$ and $M$ accessible to simulations significantly closer to those relevant for studying galaxy evolution in the allowed range of FDM masses.
\end{abstract}


\maketitle


\emph{Introduction.} Fuzzy (or wave) dark matter (FDM) is a class of ultra-light bosonic dark matter models giving rise to pronounced wave-like effects in collapsed cosmological structures \cite{Hu2000,Schive2014,Marsh2016,Niemeyer2020}. It is represented by a classical field theory for ultra-light (pseudo)scalar particles, including axion-like particles, with negligible non-gravitational interactions that reside in very low, highly populated momentum states. In plausible scenarios motivated by superstring cosmology \cite{Arvanitaki2010,Hui2017}, FDM particles are abundantly produced non-thermally in the early universe. Modeled as a non-relativistic coherent scalar field $\psi$ with mass $m$ their time evolution, to leading order, is governed by the comoving Schr\"odinger-Poisson (SP) equation \cite{Salehian2021}
\begin{align}
    \label{eq:1}
     i\hbar\frac{\partial\psi}{\partial t} & = -\frac{\hbar^{2}}{2ma^{2}}\nabla^{2}\psi+mV\psi\,\,,\nonumber\\
 \nabla^{2}V &= \frac{4\pi G}{a}\delta\rho\quad,\quad\rho=|\psi|^{2}\, ,
\end{align}
where $V$ denotes the gravitational potential and $a$ is the scale factor. The coherence wavelength $\lambdabar_\mathrm{dB}\sim \hbar/(mv)$ provides a characteristic length scale above which FDM behaves like cold dark matter (CDM) with respect to gravitational interactions~\cite{Widrow1993,Uhlemann2014,Veltmaat2018}, hence mirroring the successes of standard CDM on these scales. 

New, discriminating phenomena occur on length scales close to $\lambdabar_\mathrm{dB}$ and characteristic times $\sim \hbar/(mv^2)$. Simulations found that FDM halos host solitonic cores surrounded by a fluctuating, granular structure formed by wave interference \cite{Schive2014,Mocz2017,Veltmaat2018}. Later simulations started to include baryons~\cite{Veltmaat2020,Mocz2019}. Core formation~\cite{Levkov2018}, evolution~\cite{Veltmaat2018,DuttaChowdhury2021}, and mergers~\cite{Schwabe2016} have been further investigated. The fluctuating granules produce gravitational relaxation effects on star clusters or black holes \cite{Hui2017,Baror2019,Elzant2019,Lancaster2020} leading to strong constraints on the allowed FDM mass range~\cite{Marsh2019}. Together with bounds on suppression of small-scale power from the Lyman-$\alpha$ forest flux power spectrum \cite{Armengaud2017,Irsic2017} and the high-redshift galaxy luminosity function \cite{Schive2016,Bozek2015,Menci2017,Corasaniti2017} , they indicate a lower bound on the FDM mass of $m \gtrsim 10^{-21}$ eV. For a comprehensive review see Ref.~\cite{Niemeyer2020}.

\begin{figure*}
    \centering
    \includegraphics[width=\linewidth]{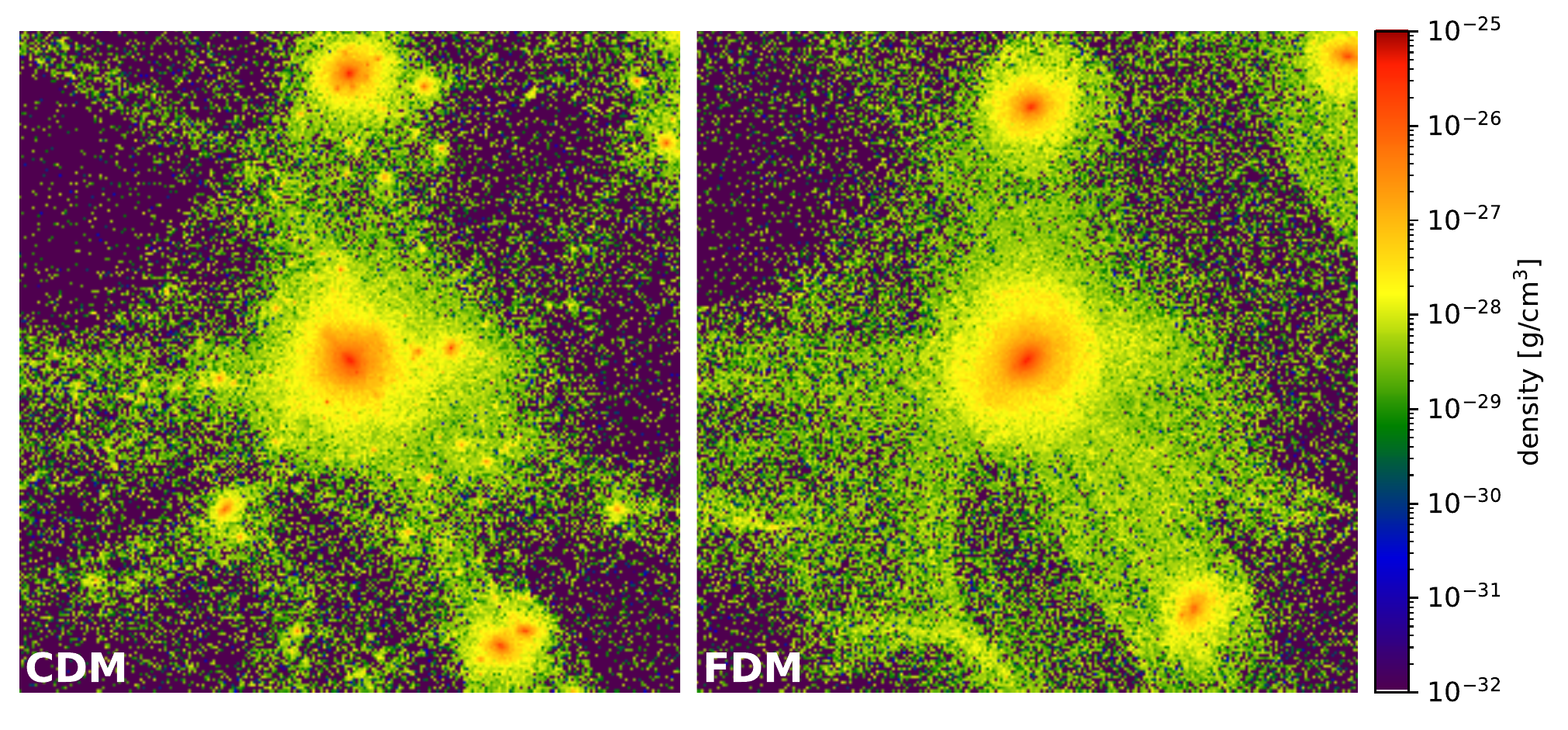}
    \caption{N-body densities at $z=0$ for CDM (left) and FDM (right) initial conditions of the \textsc{Agora} proof-of-principle halo with virial mass $M \simeq 1.7 \times 10^{11} M_\odot$ in a $1h^{-1}$ Mpc box. While the CDM density reproduces previous \textsc{Agora} results \cite{Kim2014}, the cut-off in the FDM initial perturbation spectrum results in reduced substructure.}
    \label{fig:1}
\end{figure*}

Suppression of the linear perturbation spectrum manifests itself on scales $\gg \lambdabar_\mathrm{dB}$ and is therefore accessible to standard N-body methods. On the other hand, simulations of nonlinear wave-like effects inside collapsed structures require solutions of the SP equations. These are numerically expensive since the complex phase of the wave function has to be properly resolved in the entire simulation volume including voids~\cite{Schive2014}. Methods that solve the fluid representation of \autoref{eq:1} including a quantum pressure term can overcome these restrictions but cannot account for interference patterns emerging after multi-streaming occurs \cite{Mocz2015,Veltmaat2016,Nori2018,Hopkins2019}. This approach is therefore inadequate to study wave-like effects near $\lambdabar_\mathrm{dB}$. 

For these reasons, simulations with cosmological initial conditions and statistically meaningful volumes aiming at resolving wave-like dynamics in FDM halos have been restricted to FDM masses $m\lesssim 10^{-22}$ eV, i.e. significantly below the bound from large-scale structure probes, and halo masses $M \lesssim 10^{10} M_\odot$. There is an urgent need to push computational capabilities toward higher $m$ and $M$ with realistic initial conditions. Here, we present simulations of an isolated halo from initial conditions provided by the \textsc{Agora} galaxy evolution project with final mass $M \simeq 1.7 \times 10^{11} M_\odot$, adapted for an FDM linear power spectrum with  $m = 2.5 \times 10^{-22}$ eV. The numerical resolution needed to observe the formation of a central soliton was achieved by running deep zoom-in simulations with a hybrid N-body-Schr\"odinger scheme which is a significantly improved version of the method described in Ref. \cite{Veltmaat2018}.

In Ref.~\cite{Veltmaat2018}, we combined the efficiency of N-body simulations with the accuracy of finite-difference solvers for the Schrödinger equation using adaptive mesh refinement (AMR). We conducted zoom-in simulations focusing on the inner dynamics of a few pre-selected halos. Evolving most of the simulation volume using N-body particles to accurately compute the large-scale gravitational field and the mass accretion onto the halos, the highly resolved halos themselves were evolved by explicitly solving the SP equations. The critical part was the reconstruction of the wave function from particle information at the N-body-Schr\"odinger boundaries, for which the classical wave approximation \cite{Trahan2005} was used. These simulations enabled us to investigate the dynamics of the halos' granular structure and their central solitonic cores.

The main downside of the classical wave approximation is its inability to capture the interference pattern in multi-streaming regions. The wave function therefore needs to be reconstructed in the Lagrangian volume of the halo before the onset of collapse, restricting the analysis in \cite{Veltmaat2018} to dwarf sized halos.

In this work, we present a new reconstruction method for the N-body-Schr\"odinger boundaries that fully captures the non-linear wave dynamics on a statistical level. It is  implemented in our \textsc{AxioNyx} code specialized for axion-like particle dark matter simulations \cite{Schwabe2020}. The new reconstruction scheme is closely related to the \emph{Gaussian Beam} (GB) method~\cite{Kluk1986,Kay1994,bach2002,Kay2006} originally developed for semi-classical calculations in quantum chemistry~
\cite{Ceotto2017,Patoz2018,Buchholz2018,Bertaina2019,Gabas2019}. Variations of the GB method have been used to study the dynamics around quantum barriers with discontinuous potentials~\cite{Kay2013,Jin2014,Lu2016}, including a hybrid method similar to the one described below~\cite{Jin2011}.
It also lends itself to the analysis of interacting bosons~\cite{Ray2016} and photoexcitation and photoionization~\cite{Bichkov2015}. Contrary to simple ray tracing algorithms, the GB method does not become singular at caustics~\cite{Kay1994}. See the supplementary material (appendices \ref{sec:GBM} and \ref{sec:method}) for details of the full GB method, its relation to our reconstruction scheme, and the implementation in \textsc{AxioNyx}.

Using our new GB-related technique, we can reconstruct the wave function after the pre-selected halo has already collapsed while resolving only the inner part of the halo well within its virial radius. This improvement enabled us to re-simulate the proof-of-concept test of the \textsc{Agora} High-resolution Galaxy Simulations Comparison Project \cite{Kim2014}, consisting of a dark-matter only simulation of a sub-$L_\ast$-sized galactic halo from cosmological initial conditions, with full FDM dynamics.

\begin{figure*}
    \centering
    \includegraphics[width=\linewidth]{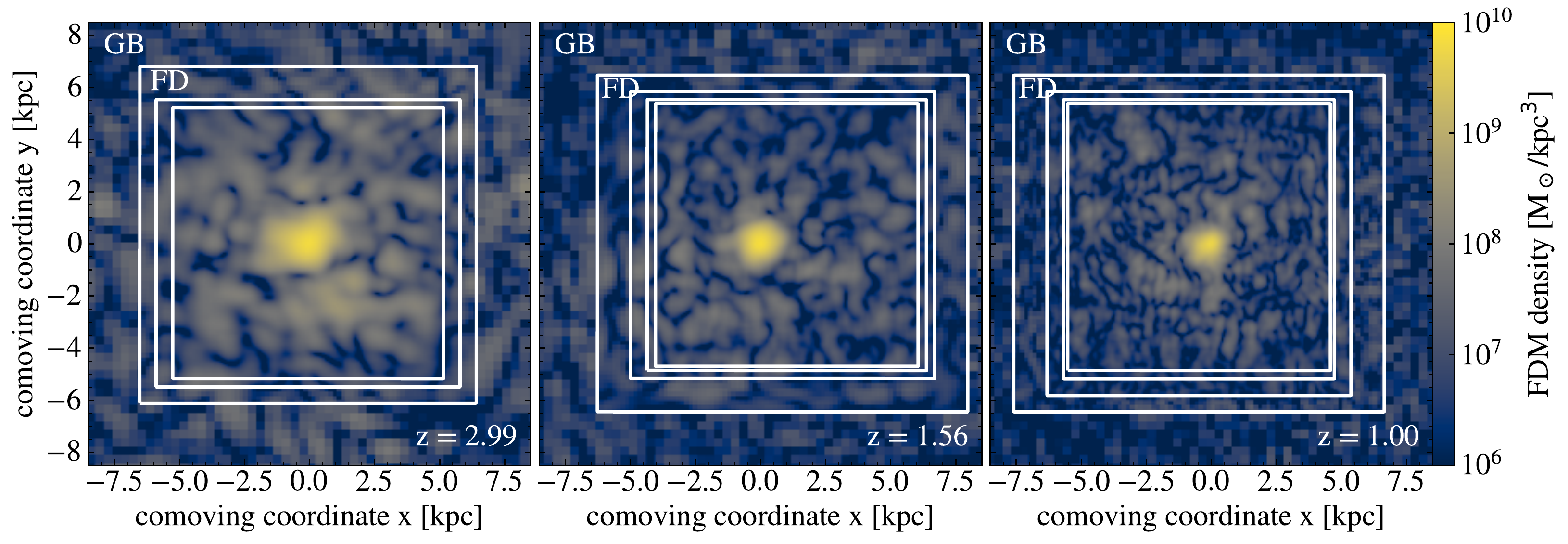}
    \caption{Density slices through the wave function at different redshifts. (left) Reconstructed wave function with self-consistent interference pattern and central solitonic core using the GB method. The wave function is evolved with a finite-difference solver on three additional levels. (middle) To ensure sufficient resolution another level is added at $z=1.56$. (right) Final snapshot at $z=1$.}
    \label{fig:2}
\end{figure*}


\emph{Simulation setup.} As part of the \textsc{Agora} project \cite{Kim2014}, several widely used cosmology codes were compared by evolving identical initial conditions in a dark matter only simulation with standard $\Lambda$CDM cosmology. Using a ($60$h${^{-1}}$ comoving Mpc)${}^{3}$ box on a $128^3$ root grid, the Lagrange patch of a pre-selected halo with virial mass $M \simeq 1.7 \times 10^{11} M_\odot$ at $z=0$ and quiescent merger history was further resolved by five static refinement levels and up to six adaptively refining levels whenever an overdensity of four or more particles was reached in a single cell. Doubling the resolution per refinement level, the simulation was thus resolved down to $326$ comoving pc. In order to properly resolve the FDM interference patterns of the pre-selected halo in our simulations, the wave function is reconstructed on higher levels reaching a final resolution of $20$ comoving pc.

The pre-selected halo is first re-run in pure N-body mode with \textsc{AxioNyx} in order to ensure consistency with \cite{Kim2014}. Constructing $\Lambda$CDM initial conditions at redshift $z=100$ with \textsc{Music}\cite{Hahn2011} as specified by the \textsc{Agora} project, we recover the expected final density configuration at $z=0$. Using the publicly available analysis scripts of the \textsc{Agora} project, we obtain the same density slice plot through the halo center. The consistency of the numerical results can be seen by comparing our \autoref{fig:1} (left) with Figure 3 in \cite{Kim2014}.

FDM cosmology is characterized by a cut-off in the initial transfer function \cite{Hu2000}. We obtained the corresponding FDM transfer function using \textsc{AxionCAMB} \cite{Hlozek2015} with $m = 2.5 \times 10^{-22}$ eV.  Initial conditions were created with \textsc{Music} \cite{Hahn2011}, keeping the original large scale features but suppressing small scales. As expected, the N-body simulation conducted with \textsc{AxioNyx} reveals a final state which is effectively a smoothed version of the CDM final state. As seen in \autoref{fig:1} (right), in the FDM run only the largest halos collapsed.

Restarting the FDM simulation at redshift $z=3$, after the pre-selected halo has fully collapsed, we further refine its innermost region. After subtracting the halo's mean velocity from all particles in order to reduce resolution requirements of large phase gradients, we reconstruct the wave function on the $11$th level and add three additional levels evolved by \textsc{AxioNyx}'s finite-difference Schrödinger solver. We thus establish a hybrid method similar to the one presented in \cite{Veltmaat2018}. The important improvement here is the full reconstruction of the FDM interference pattern. 

As the resolution of the granules on the GB level deteriorates over time, interpolation to the next finer FD level results in a continuous mass increase. We compensate for it by rescaling the wave function on the coarsest FD level such that its average density coincides with the N-body density obtained from the underlying beams. At each time step, the rescaling does not exceed a per-mil level change.  


\emph{Numerical results.} \autoref{fig:2} shows density slices through the wave function at different redshifts. The first one at $z=2.99$ represents the wave function immediately after reconstruction. The characteristic interference pattern and the central solitonic core are clearly visible. As the collapsed halo decouples from the expanding background, its substructure shrinks relative the simulation box. In order to ensure sufficient resolution, we insert an additional refinement level after redshift $z = 1.56$ and stop the simulation at $z=1$.  

\begin{figure*}
    \centering
    \includegraphics[width=\linewidth]{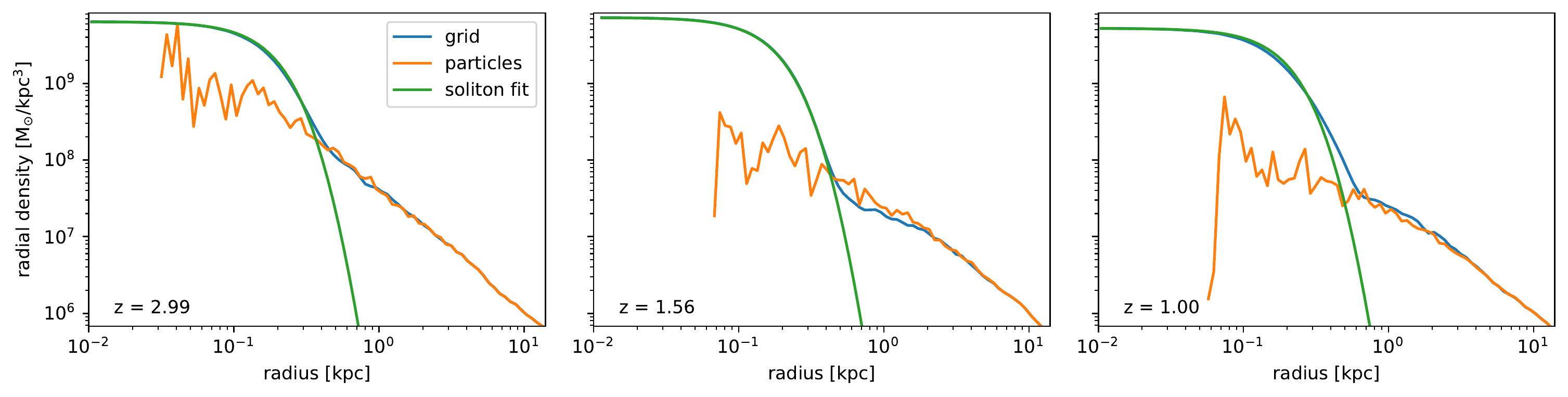}
    \caption{Radial density profiles of the FDM wave function (blue) at different redshifts. They are well fitted by a soliton profile (green) transitioning to an outer NFW-like profile indistinguishable from the one obtain utilizing the underlying N-body particle information (orange).}
    \label{fig:3}
\end{figure*}

\begin{figure*}
    \centering
    \includegraphics[width=\linewidth]{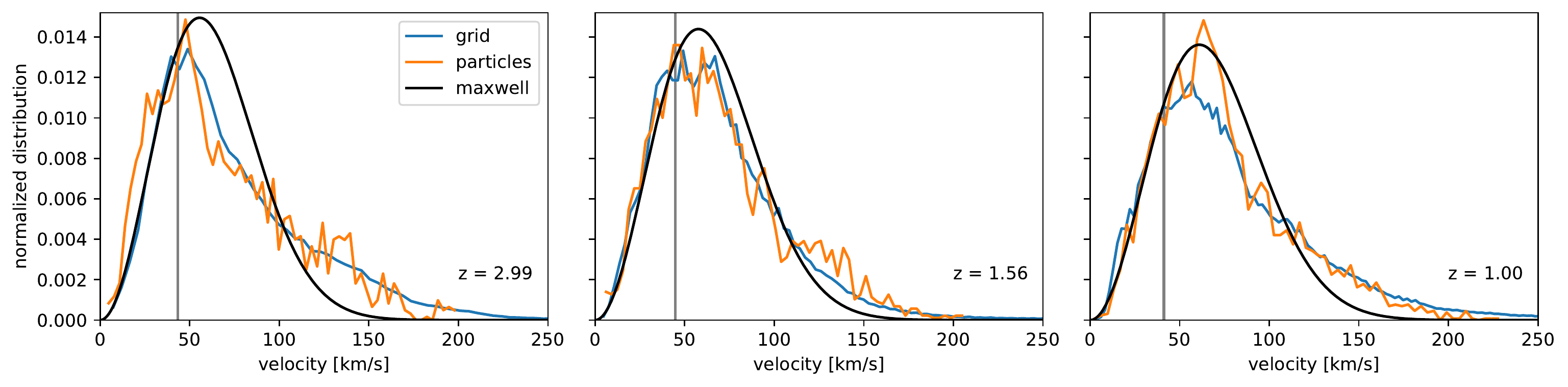}
    \caption{Velocity spectra calculated on the finest AMR level at different redshifts. The FDM wave function's spectra (blue) are comparable to the underlying particle velocity dispersions (orange) and are close to a Maxwellian distribution (black). The vertical lines mark the velocities $v_c$ from \autoref{eq:soliton_vel}.}
    \label{fig:4}
\end{figure*}

Radial density profiles centered around the halo's density maxima at different redshifts are shown in \autoref{fig:3}. The inner region is well fitted by a solitonic core profile \cite{Schive2014}
\begin{align}
    \rho_c(r) \simeq \rho_{0}\left(1 + 0.091\left(\frac{r}{r_c}\right)^2\right)^{-8}\,,
    \label{eq:soliton_profile}
\end{align}
where $r_c$ is the core radius at which the density has dropped to half its central value
\begin{align}
    \rho_{0}\simeq 3.1\times 10^{6}\left(\frac{2.5\times 10^{-22}\text{eV}}{m}\right)^{2}\left(\frac{\text{kpc}}{r_{c}}\right)^{4}\;\frac{M_{\odot}}{\text{kpc}^{3}}\, .
\end{align}
The outer angular-averaged density profile is NFW-like and statistically indistinguishable from CDM, confirming previous results. 

\begin{figure*}
    \centering
    \includegraphics[width=\linewidth]{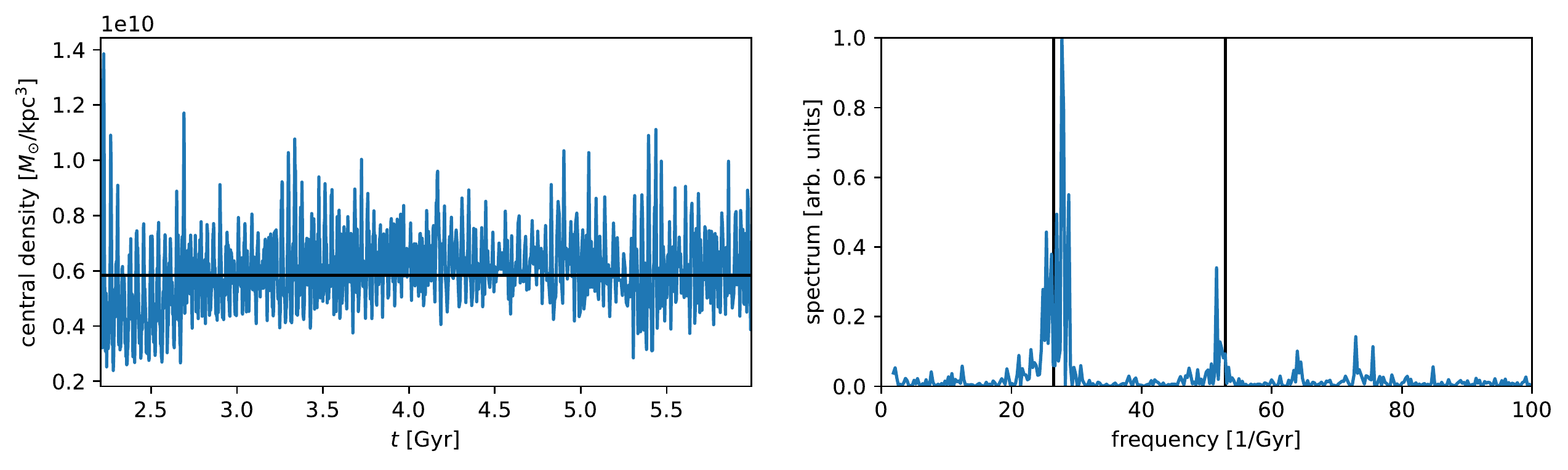}
    \caption{(left) The central soliton density oscillates around an average value indicated by the black line. (right) Using \autoref{eq:freq} it defines a quasi-normal soliton frequency and its first higher harmonic (vertical black lines), which are both well matched by the numerically obtained frequency spectrum of the oscillating central density.}
    \label{fig:5}
\end{figure*}

The wave function's velocity distribution is calculated on the finest level~\cite{Veltmaat2018}:
\begin{align}
    f(\mathbf{v}) = \frac{1}{N}\abs{\int \mathrm{d}^3\mathbf{x}\exp\left(-im\mathbf{v}\cdot\mathbf{x}/\hbar\right)\psi(\mathbf{x})}^2\,,
\end{align}
with normalization factor $N$. It was previously shown to match the underlying particle velocity distribution~\cite{Veltmaat2018}, consistent with the equivalence of the coarse-grained Schrödinger and Vlasov equations known as the Schrödinger-Vlasov correspondence \cite{Widrow1993,Uhlemann2014,Mocz2018}. As seen in \autoref{fig:4}, we recover the same similarity to the beams' velocity distribution on the finest level proving the consistency of the reconstructed interference pattern on a statistical level. Both spectra are fitted well by a Maxwellian distribution
\begin{align}
    f_M(v)\mathrm{d}v = 3\left(\frac{6}{\pi}\right)^{1/2}\frac{v^2}{v_0^3}\exp\left(-\frac{3}{2}\frac{v^2}{v_0^2}\right)\mathrm{d}v\,,
    \label{eq:maxwellian}
\end{align}
with free parameter $v_0$. 

It was found numerically that solitonic core radii are correlated with the peaks of the velocity distributions~\cite{Mocz2017,Schwabe2020}:
\begin{align} 
    v_c = \frac{2\pi}{7.5}\frac{\hbar}{mr_c}\,.
    \label{eq:soliton_vel}
\end{align}
For virialized halos, this implies that the growth of solitons is suppressed once their virial temperature reaches that of their host halo \cite{Eggemeier2019,Niemeyer2020,Chen2020}. The velocities corresponding to $v_c$ at different redshifts are displayed as vertical lines in \autoref{fig:4}.

On average, the solitonic core's central density, depicted in \autoref{fig:5}, remains constant over the entire simulation period and oscillates on the quasi-normal frequency of the excited soliton~\cite{Guzman2004,Veltmaat2018}
\begin{align}
 f = 10.94 \left(\frac{\rho_c}{10^9 \,\text{M}_\odot \text{kpc}^{-3}}\right)^{1/2} \text{Gyr}^{-1}\,.
\label{eq:freq}
\end{align}


\emph{Conclusions.} We presented the largest FDM cosmology zoom-in simulation to date with an effective resolution of $4.2\times 10^6$ cells in all three spatial dimensions, exceeding previous FDM cosmology simulations by roughly six orders of magnitude in the number of effective grid points~\cite{Schive2014,Veltmaat2018,May2021}. The spatial resolution at the highest refinement level of $20$ comoving pc allowed the full wave-like simulation of a sub-$L_\ast$-sized galactic halo from cosmological initial conditions with FDM mass $m = 2.5 \times 10^{-22}$ eV. 

In order to facilitate comparison with CDM, we chose to re-run the proof-of-concept dark-matter only simulation of the \textsc{Agora} code comparison project \cite{Kim2014}. We assumed standard $\Lambda$CDM cosmology in a ($60$h${^{-1}}$ comoving Mpc)${}^{3}$ box with FDM initial conditions for AMR simulations of a single selected halo. Using a $128^3$ root grid, the halo's Lagrange patch was further resolved by five static refinement levels and up to ten adaptively refining levels.

This deep zoom-in simulation was made possible by extending the \textsc{AxioNyx} code with an improved hybrid N-body-Schr\"odinger method building on the technique used in Refs. \cite{Veltmaat2018, Veltmaat2020}. Its key advantage over full Schrödinger-Poisson simulations is the capability to solve the Schrödinger equation only in highly refined subvolumes, while relying on the Schrödinger-Vlasov correspondence to treat dark matter in regions with coarser refinement level as N-body particles. This approach drastically reduces the required spatial resolution in most of the computational volume. 

The main improvement over previous versions is a new reconstruction scheme at the N-body-Schrödinger boundaries based on a simplified version of the Gaussian Beam method for solving the Schrödinger equation, providing the statistically correct reconstruction of the solitonic core and the interference pattern in the central region of the collapsed halo. This methods requires only the complex phase co-evolved with each N-body particle to reconstruct an FDM wave function in fully non-linear density fields. The wave-particle conversion can therefore begin after the halo has collapsed and be confined to a region well within its virial radius.

We recover the radial FDM density profiles with a solitonic core embedded in a fluctuating halo whose averaged density profile is consistent with an NFW behavior. The FDM wave function's velocity spectrum on the finest AMR level coincides with the underlying particle velocity dispersion and resembles a Maxwellian distribution. It peaks close to the solitons virial velocity implying that the core is in kinetic equilibirum with its surrounding. The soliton is in an exited state dominated by the quasi-normal frequency mode. 

Our result is a proof-of-concept demonstration of the hybrid N-body-Schrödinger method, pushing the range of $m$ and $M$ accessible to simulations significantly closer to those relevant for studying galaxy evolution in the allowed range of FDM masses $m \gtrsim 10^{-21}$ eV. Future simulations will need to include the effects of baryons and star formation whose strong impact on the core soliton was shown in Ref. \cite{Veltmaat2020}. We also expect that it will enable simulations of gravitational relaxation and heating in FDM halos from realistic initial conditions and a variety of merger histories.  

Beyond FDM research, the nonlinear dynamics of the Schrödinger-Poisson equation is relevant for studying the gravitional fragmention of the inflaton field in scenarios with early matter domination \cite{Musoke2020,Eggemeier2021} including the formation of ``inflaton stars'' \cite{Niemeyer2020b,Eggemeier2021b}, as well as axion miniclusters and axion stars formed from QCD axion dark matter \cite{Eggemeier2019,Eggemeier2020}. These simulations can equally benefit from the method presented here.


\emph{Acknowledgements.} We thank Benedikt Eggemeier, Mateja Gosenca, and Richard Easther for important discussions. Computations described in this work were performed with resources provided by the North-German Supercomputing Alliance (HLRN). We acknowledge the yt toolkit~\cite{Turk2011} that was used for the analysis of numerical data. BS acknowledges support by the Deutsche Forschungsgemeinschaft and by grant PGC2018-095328-B-I00(FEDER/Agencia estatal de investigaci\'on). 

\bibliography{semiclassical_FDM}

\appendix

\section{Gaussian Beam Method}
\label{sec:GBM}
\subsection{Full Gaussian Beam Method}

Any wave function $\braket{x | \psi}\equiv\psi(x)\in L^{2}(\mathbb{R}^{d},\mathbb{C})$ can be decomposed into coherent Gaussian wave packets
\begin{align}
\label{eq:2}
    \braket{x | p,q,\gamma,0}\equiv&\det\left(\frac{2 \text{Re}\,\gamma}{\pi}\right)^{1/4}\exp\text{[}ip\cdot(x-q)/\hbar\nonumber\\
    &-(x-q)^{T}\gamma(x-q)\text{]}\, ,
\end{align}
via the Fourier-Bros-Iagolnitzer (FBI) transformation $T_{\gamma}:L^{2}(\mathbb{R}^{d},\mathbb{C})\mapsto L^{2}(T^{*}\mathbb{R}^{d},\mathbb{C})$:
\begin{align}
    \label{eq:3}
    T_{\gamma}[\psi](q,p)&\equiv\frac{1}{(2\pi\hbar)^{d/2}}\int_{\mathbb{R}^{d}}\braket{p,q,\gamma,0 | x} \braket{x | \psi}\text{d}x\nonumber\\
    &\equiv\braket{p,q,\gamma,0 | \psi}
\end{align}
and its inverse $T^{*}_{\gamma}:L^{2}(T^{*}\mathbb{R}^{d},\mathbb{C})\mapsto L^{2}(\mathbb{R}^{d},\mathbb{C})$:
\begin{align}
    \label{eq:4}
    T^{*}_{\gamma}[u](x)\equiv\frac{1}{(2\pi\hbar)^{d/2}}\int_{T^{*}\mathbb{R}^{d}} \braket{x | p,q,\gamma,0}\braket{p,q,\gamma,0 | u}\text{d}p\text{d}q
\end{align}
since \cite{bach2002}
\begin{align}
    \label{eq:5}
    \braket{x | \psi} = \braket{x | T_{\gamma}^{*}T_{\gamma}\psi} \, .
\end{align}
Given any initial wave function $\psi$ we can thus populate the phase-space $T^{*}\mathbb{R}^{d}$ with coherent Gaussian beams $\braket{x | p,q,\gamma,0}$ weighted by $(2\pi\hbar)^{-d/2}\braket{p,q,\gamma,0 |\psi}$. \autoref{eq:5} then implies that we can recover $\psi$ as the integral over all beams.

Here, $\gamma\in\mathbb{C}^{n\times n}$ is symmetric and its real part is positive definite. Its imaginary part can be used to extend the Taylor expansion in the initial phase reconstruction to second order.

The FBI transformation is identical to the Husimi-Q representation upto a different overall phase definition that cancels out when calculating the density $\rho = |\psi|^{2}$ \cite{Uhlemann2014}.

Instead of directly evolving $\psi$ we can now update each beam separately. Using the FBI-transformation, it was shown in \cite{Kay1994} that $\psi$ is recovered at any given time by integrating over all time evolved beams. Thus,
\begin{align}
    \label{eq:6}
    \psi(x,t) =& \frac{1}{(2\pi\hbar^{\prime})^{d}}\int_{T^{*}\mathbb{R}^{d}} \braket{x | p'_{t},q_{t},\gamma_{t},t}\nonumber\\
    &\times\int_{\mathbb{R}^{d}}\braket{p'_{0},q_{0},\gamma_{0},0 | y} \braket{y | \psi_{0}}\text{d}y\text{d}p'_{0}\text{d}q_{0}\, ,
\end{align}
with initial conditions $\psi(x,0)\equiv\psi_{0}$ and time-dependent Gaussian beams
\begin{align}
    \label{eq:7}
    \braket{x | p'_{t},q_{t},\gamma_{t},t}\equiv C_{pqt}e^{iS'_{pqt}/\hbar'}\braket{x | p'_{t},q_{t},\gamma_{t},0}\, .
\end{align}
Given the classical Hamiltonian function corresponding to \autoref{eq:1}
\begin{align}
    \label{eq:8}
    H'(p_{t},q_{t})  = \frac{p'_{t}{}^{2}}{2a^{2}}+V(q_{t})\, ,
\end{align}
the beams move on their classical trajectories
\begin{align}
  \label{eq:9}
   \frac{dq_{t}}{dt} =& \frac{\partial H'}{\partial p'_{t}} = \,\frac{p'_{t}}{a^{2}} \quad , \quad 
   \frac{dp'_{t}}{dt} = -\frac{\partial H'}{\partial q_{t}} = - \frac{dV}{dq_{t}}\, ,
\end{align}
where primes denote mass scaled quantities (e.g. $p'=p/m$). We introduce primed quantities in order to avoid an explicit mass dependence in the Schr\"odinger equation which does not depend separately on $\hbar$ and $m$, but only on their ratio $\hbar'$. 

As for ray tracing methods, the time evolution of the central phase $S_{pqt}$, being the action of the system, is governed by its Lagrangian $L$:
\begin{align}
  \label{eq:10}
  S'_{pqt} = \int_{0}^{t}L'\text{d}t =\, \int_{0}^{t}\left[\frac{p'_{t}{}^{2}}{2a^{2}}-V(q_{t})\right]\text{d}t\,. 
\end{align}
The pre-factor $C_{pqt}$ accounts for the time varying Jacobian of the system. It is given by
\begin{align}
    \label{eq:11}
    C_{pqt}& = \det\left(\frac{\pi}{2 \text{Re}\,\gamma_{t}}\right)^{1/4}\det\left(\frac{\pi}{2 \text{Re}\,\gamma_{0}}\right)^{1/4}\left(\frac{1}{\pi}\right)^{d/2}\nonumber\\
    \times&\det\left[\gamma_{t}\frac{\partial q_{t}}{\partial q_{0}}-\frac{1}{2i\hbar'}\frac{\partial p'_{t}}{\partial q_{0}}+\left(\frac{\partial p'_{t}}{\partial p'_{0}}-2i\hbar'\gamma_{t}\frac{\partial q_{t}}{\partial p'_{0}}\right)\gamma_{0}^{*}\right]^{1/2}
\end{align}
with initial Jacobian matrices
\begin{align}
    \label{eq:12}
    J_{0} = \begin{pmatrix} \partial q_{0}/\partial q_{0} & \partial q_{0}/\partial p'_{0} \\
    \partial p'_{0}/\partial q_{0} & \partial p'_{0}/\partial p'_{0}\end{pmatrix} = \begin{pmatrix} \mathds{1} & 0 \\
    0 & \mathds{1}\end{pmatrix}\,\, .
\end{align}
Their time evolution is obtained by differentiating \autoref{eq:9}:
\begin{align}
    \label{eq:12b}
    \frac{d J_{t}}{dt} = UJ_{t}\quad , \quad U = \begin{pmatrix} 0 & a^{-2}\\
    -\frac{d^{2}V}{dq_{t}^{2}} & 0 \end{pmatrix}\,\, .
\end{align}
The matrix $\gamma_{t}$ can be arbitrarily chosen as long as it fulfills the requirements of $\gamma$, changes continuously with time and stays finite. Since $\gamma_{t}=\gamma_{0}$ initially, we have $C_{pq0}=1$, $S'_{pq0}=0$ and \autoref{eq:6} reduces to \autoref{eq:5} as required.

\subsection{WKB-like Gaussian Beam Method}
\label{sec:WKB}

The GB method presented in the previous section requires the sampling of the six dimensional phase space, which is numerically not feasible for cosmological simulations. In this section we therefore develop a WKB-like approximation to the full GB method.

Assuming WKB initial data
\begin{align}
    \label{eq:25}
    \psi_{0}(y) \simeq A_{0}(q_{0})\exp[i(S'_{0}(q_{0})+\overline{p}'(q_{0})(y-q_{0}))/\hbar']
\end{align}
the beam weights are given by \cite{Widrow1993}
\begin{align}
    \label{eq:26}
    (2\pi\hbar')^{-d/2}&\braket{p'_{0},q_{0},\gamma_{0},0 |\psi_{0}}\nonumber\\
    =& \det\left(\frac{2 \text{Re}\,\gamma_{0}}{\pi}\right)^{1/4}A_{0}\exp[iS'_{0}(q_{0})/\hbar']f(p'_{0}|\overline{p}',\sigma_{p'})
\end{align}
with normal distribution
\begin{align}
    \label{eq:27}
    f(p'_{0}|\overline{p}',\sigma_{p'}) =& \det\left(\frac{1}{4\pi\hbar'^{2}\text{Re}\gamma_{0}}\right)^{1/2}\nonumber\\
    &\times\exp\left[-(p'_{0}-\overline{p}')\gamma_{0}^{-1}(p'_{0}-\overline{p}')/(4\hbar'^{2})\right]\,\,. 
\end{align}
Thus, the convolution of $\psi_{0}(y)$ with $\braket{y | p'_{0},q_{0},\gamma_{0},0}$ localizes the wave function around $q_{0}$ with standard deviation $\sigma_{q}=1/\sqrt{2\text{Re}\gamma_{0}}$ (cf. \autoref{eq:3}) and around $\overline{p}'$ with $\sigma_{p'}=\hbar'\sqrt{2\text{Re}\gamma_{0}}$. This ensures the uncertainty principle $\sigma_{q}\sigma_{p'}=\hbar'$. Note that the usual factor of one half enters when squaring the amplitude in order to obtain the density. 

Since $\gamma_{0}$ is a free parameter, we can formally take it to be zero. In this limit, \autoref{eq:27} becomes a delta distribution $\delta(p'_{0}-\overline{p}'(q_{0}))$. It can be used to get rid of the momentum integral in \autoref{eq:6}. Additionally, the term proportional to $\gamma_{0}$ vanishes in \autoref{eq:11}. We thus only need to evolve half of the Jacobian matrix (\autoref{eq:12}). 

\subsection{Fixed amplitude GB method}
\label{sec:fagbm}

The WKB-like method is not stable enough for a realistic reconstruction of fully collapsed FDM halos. The problem is the extreme deformation of phase space in those regions resulting in strongly fluctuating summands in \autoref{eq:11} due to the multiplication of numbers of very different size and changing sign. The reconstruction can then lead to densities that are orders of magnitude higher than expected from analogous CDM runs and mass conservation can not be ensured.

In order to overcome this problem the beams' amplitudes can be fixed to their initial values
\begin{align}
    \label{eq:32}
    C_{pqt} = \frac{A_{0}\gamma_{t}^{3/2}\Delta^{3}q_{0}}{\pi^{3/2}}\, .
\end{align} 
On the one hand this makes the algorithm much more efficient as we only need to store and evolve the complex phase but not the Jacobian of each beam. On the other hand the wave function as reconstructed with such an algorithm will only be statistically equivalent to the actual wave function. Our aim is therefore to use the particle information in the center of a pre-selected halo to construct a wave function with the statistically correct interference pattern as quantified e.g. in \cite{Veltmaat2018}.

The full GB method, like the Schr\"oedinger-Vlasov correspondence, is applicable when higher than second order spacial derivatives of the potential are negligible. Cosmological simulations of FDM have confirmed the applicability of the latter. From \autoref{eq:11} and \autoref{eq:12b} we see that fixing the amplitude means neglecting second order derivatives as well. Since they are associated with velocity dispersion in halos and filaments, it is serendipitous that these contributions cancel out statistically as we show by numerically proving the validity of the fixed amplitude GB method in FDM cosmology simulations. Note though, that the fixed amplitude GB method is very similar to the single-Hessian thawed Gaussian approximation which was also shown to produce statistically correct results in semi-classical calculations of molecular spectra~\cite{Begusic2019}.

The fixed amplitude GB method is also similar to the previously used classical wave approximation \cite{Veltmaat2018}. Besides a slightly different kernel, the main difference is that here we do not use \autoref{eq:6} just for the wave function's phase while keeping the CDM-like N-body density, but utilize the full reconstructed wave function.

\section{Implementation in AxioNyx}
\label{sec:method}

We implemented the fixed amplitude GB method presented in appendix \ref{sec:fagbm} within the publicly available cosmology code \textsc{AxioNyx} as a new particle container storing not only standard N-body information like positions $q$, velocities $v$, and masses $M$, but also phases $S'$ for all particles. We will call N-body particles with additional phase information \textit{beams}. We initialize one beam on every initial grid cell center. Refined regions thus start with a higher beam number density. As in \cite{Veltmaat2018} the initial beam phases $S'_{0}$ are reconstructed employing \textsc{AxioNyx}'s Poisson solver for integrating
\begin{align}
   \nabla^{2} S'_{0}(x) = a\nabla\cdot v(x)\,\, .
\end{align}
Their time evolution is given by
\begin{align}
  \frac{\text{d}S'}{\text{d}t} = \frac{v^{2}}{2}-V(q)\,. 
\end{align}
The FDM wave function can be recovered at any time from the information stored in the time evolved beams by summing over them:
\begin{align}
    \psi&(x,t) = \sum_{\text{beams}}W(x-q)\exp[i(S'+v\cdot (x-q)a/\hbar']\, ,\nonumber
\end{align}
with Gaussian kernel
\begin{align}
    W(x-q) = \left(\frac{M}{(\Delta x)^{3}}\right)^{1/2}\left(\frac{2\gamma}{\pi}\right)^{3/4}\exp[-\gamma(x-q)^{2}/(\Delta x)^{2}]\, .
\end{align}
We choose $\gamma=1/32$ and truncate the kernels at three standard deviations. The width of the kernel depends on the spacing $\Delta x$ of the grid onto which the wave function in reconstructed. We checked that varying $\gamma$ by fifty percent did not change our results qualitatively.

In addition to the preexisting time step criteria enforcing the CFL condition for the beams and the cosmological constraint that limits the time step so that the simulated universe only expands by some fractional amount, we require the change of the beams' phases to be at most a fraction $\kappa_{\text{beam}}=0.5$ of $2\pi$:
\begin{align}
  \label{eq:13}
    \Delta t_{\text{beam,kin}} =& \text{min}\left(\kappa_{\text{beam}}\frac{4\pi\hbar'} {v^{2}}\right)_{L}\, ,\nonumber\\
    \Delta t_{\text{beam,pot}} =& \text{min}\left(\kappa_{\text{beam}}\frac{2\pi\hbar'}{V}\right)_{L}\, ,
\end{align}  
where the potential $V$ is evaluated at the beams' centers. The subscript $L$ emphasizes that the minimum is taken over all beams on a given level $L$.

\end{document}